\documentclass[aps,showpacs,twocolumn,amsmath,amssymb,prl]{revtex4-1} 
\usepackage{graphicx} 
\usepackage{bm} 
\usepackage{color}

\begin{document} 
 
\title{Nonlinear transmission matrices of random optical media}
 \author{A. Fleming$^1$, C. Conti$^{2\dag}$, A. Di Falco$^1$} 
\email{adf10@st-andrews.ac.uk, \dag claudio.conti@uniroma1.it} 
%\thanks{Corresponding author}
\affiliation{
$^1$School of Physics and Astronomy, University of St. Andrews, North Haugh, St. Andrews KY16 9SS, UK\\
$^2$ Department of Physics, University Sapienza, Piazzale Aldo Moro 2, 00185, Rome, Italy}

%\date{\today} 

%\pacs{05.45.Mt, 42.25.-p, 05.40.-a} 
 
\begin{abstract}
Random media with tailored optical properties are attracting burgeoning interest for 
applications in imaging, biophysics, energy, nanomedicine, spectroscopy, cryptography and
telecommunications. 
A key paradigm for devices based on this class of materials is the transmission matrix, the tensorial link
between the input and the output signals, that describes in full their optical behavior.
The transmission matrix has specific  statistical properties, as the existence of lossless channels, 
that can be used to transmit information, and are determined by the disorder distribution.
In nonlinear materials, these channels may be modulated and the transmission matrix tuned accordingly.
Here we report the direct measurement of the nonlinear transmission matrix of complex materials, exploiting the strong optothermal nonlinearity of scattering Silica Aerogel (SA).
We show that the dephasing effects due to nonlinearity are both controllable and reversible,
opening the road to applications based on the nonlinear response of random media.
\end{abstract}
\maketitle

The transmission of light in the presence of diffusion and multiple scattering can be effectively controlled manipulating its wavelength, polarization and spatio-temporal dynamics 
\cite{van2011frequency,kohlgraf2010transmission,guan2012polarization,lemoult2009manipulating}.
This rich behavior is enabled by a large multitude of optical modes that can interact during propagation, resulting in interest from a variety of fields \cite{Mosk2012, gibson2005recent, rockstuhl2010comparison,Riboli_NatMat14,Uppu2018}. 
A proper shaping of the beam wave-front is key for engineering the propagation of light in scattering media, exciting specific transmissive channels, which
route information and energy in otherwise diffusive materials and devices
\cite{kim2013relation, rotter2017light}.

Typically, this approach is based on iterative algorithms that modify the state of the light at the input of the random material, 
until a pre-determined figure of merit is obtained at the output. This brute-force approach can be implemented very efficiently 
and it has enabled a number of breakthroughs in the field, such as diffraction limit beating devices \cite{Vellekoop2010exploiting,van2011scattering}, 
optical tweezing in biological media \cite{vcivzmar2010situ}, and focusing of light in time and space \cite{aulbach2011control}. 
The merit of this technique is to treat the random material as a black box, 
in which light is efficiently coupled to and transported by transmission channels created by long-range intensity 
correlations as a result of interference effects \cite{van1990observation}.

The ability to address individual transmission channels has facilitated the development of direct measurement techniques 
to experimentally acquire the transmission matrix (TM) of the sample \cite{popoff2010}. 
This matrix contains the full information of the optical channels in the medium 
and permits to address their interaction as they transport light through the material. 
The knowledge of the TM enables either to directly shape the input beam to obtain a desired pre-designed output, or to retrieve the unknown input that generates a given output.
This powerful technique has been extensively adopted to create remarkable optical devices \cite{guan2012polarization,popoff2010image,redding2013compact,vcivzmar2012exploiting}.

The possibility of controlling complex light-matter interaction is not limited to linear propagation, but also include nonlinear optics.
The effect of nonlinearity in three-dimensional nonlinear media has been considered in ab-initio large scale simulations \cite{Conti07},
and  pioneering experiments revealing evidence of the nonlinear phase-shift of transmission channels were realized \cite{abb2011ultrafast},  
including waveform shaping \cite{katz2011focusing,Mosk2012,frostig2017focusing}. 
However, to date, no direct measurements of the transmission matrix of nonlinear random media has been reported.

In the following, we focus on the opto-thermal nonlinearity in SA. 
With a series of pump-probe experiments, to give a phenomenological demonstration that nonlinear effects are significant, we first show that - similarly to the case of ultrafast nonlinearity \cite{abb2011ultrafast} - the coupling to the transmission channels can be dynamically tuned because of the thermal nonlinearity. We then characterize the TM of the sample in nonlinear regime. 
We hence quantify the effect of nonlinearity in modifying the transmission channels and demonstrate the all-optical control of light propagation in random media.\\
\begin{figure}[htbp]
\centering
\includegraphics[width=8.6 cm]{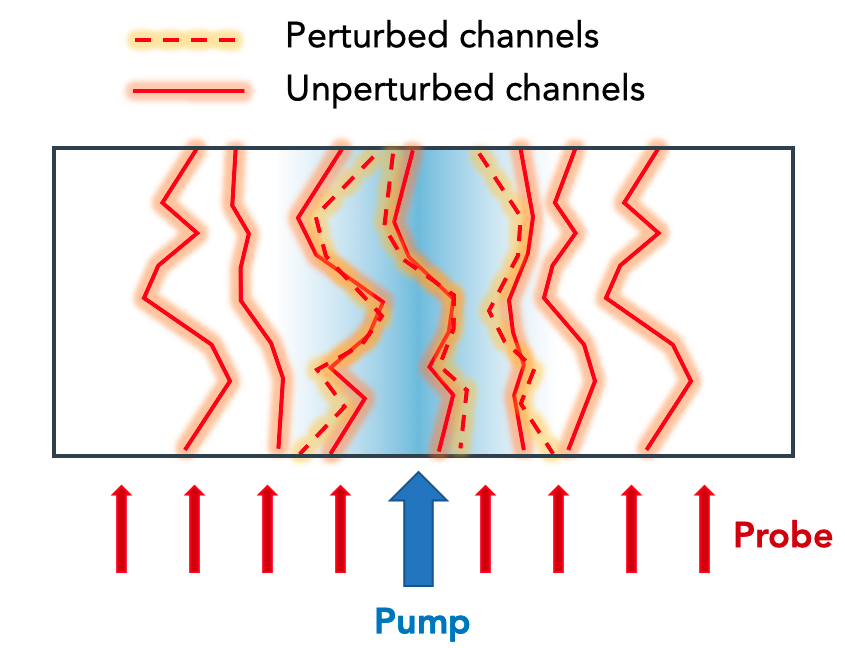}
\caption{Sketch of the formation dynamics of transmissive channels in a pump/probe configuration.}
\label{fig1}
\end{figure}

The TM is a matrix of complex coefficients ${k_{mn}}$, that relates the $n^{th}$ input fields and  $m^{th}$ output fields.
In this way the output of the $m^{th}$ mode, $E_m^{out}$ is given by the sum of input modes multiplied by the $m^{th}$ row of the TM:

 \begin{equation}
E_{m}^{out}=\sum_n k_{mn} E_n^{in}.
 \end{equation}

 The linear transmission matrix is denoted in the following as $K_{off}$, corresponding to the matrix measured when the pump beam is off.

%{\bf ------------- claudio starts here} 

%We recall that in the absence of any optimization
%$\langle |k_{mn}|^2\rangle=~1$, and this holds true
%for the transmission matrix in the presence of the perturbation
%\begin{equation}  
%  \langle |k_{mn}^{\text{NL}}|^2 \rangle= \langle |k_{mn}|^2\rangle.    
%\end{equation}  

Following the arguments developed in \cite{vellekoop2008phase}, the perturbed transmission matrix element can then be written as    
  \begin{equation}
    k_{mn}^{\text{NL}}=k_{mn}\frac{1+\xi_{mn}}{\sqrt{1+2\phi_{NL}^2}}
    \simeq k_{mn}e^{\imath \kappa_{mn}\phi_{NL} },   
    \label{eq:kprime1}
  \end{equation}
  where $\xi_{mn}$ is a complex Gaussian variable with zero mean and (for small perturbations $\phi_{NL}$) $\langle |\xi_{mn}|^2 \rangle=2\phi_{NL}^2$.
  For the modal dependent coefficients it holds
  \begin{equation}
    \kappa_{mn}\phi_{NL}\simeq \Im({\xi_{mn}}),
    \end{equation}
    such that $\phi_{NL}$  represents the average phase shift of the mode
   with $\langle |\kappa_{mn}|^2 \rangle=1$.
  As derived in details in the supplementary material, one can write
      \begin{equation}
      \phi_{NL} \simeq  \frac{\pi \omega}{2}
      \sqrt{\langle |\int \Delta\varepsilon({\bf r}) \rho({\bf r},\omega)
      \mathrm{d} {\bf r}|^2 \rangle}     
  \end{equation}

where $\Delta \varepsilon({\bf r})$ is the relative permittivity perturbation and $\rho(r,\omega)$ is the local density of states.
Hence those modes that are mostly overlapped with the index perturbation are subject to a phase-shift.
For a thermo-optical nonlinearity, letting $I({\bf r})$ the pump intensity that induces a temperature profile $\Delta T({\bf r})$, one has
$\Delta \varepsilon=\Delta T \partial \varepsilon/\partial T$.

%\adf{[CUT ALL?] For a focusing experiment, one can estimate $\phi_{NL}$ following the arguments
%of {\bf VELLEKOOP}: the maximum intensity drops
%by a factor of the order of $\exp{(-\phi_{NL}^2)}$, which can be used the experiments and estimated $\phi_{NL}$ which is proportional to the pump power.}

%{\bf ------------ claudio ends here }

In terms of a scattering process, one can visualize the process as sketched in figure \ref{fig1}. 
The spatial modulation of the input beam optimizes the transmission of the channels in the random media that gain a specific phase, e.g. to produce a focalized spot at the output. 
The nonlinear index perturbation induces multiple novel channels, with amplitude and phase that depend on the power of the pump beam, that also interfere at the output. 
This additional channels may be explained in terms of the perturbed Green function, as detailed in the SI.
For increasing pump power, the additional paths tends to phase-shift with respect to the unperturbed path. 
%This nonlinear phase-shift causes the blurring of the disordered lens focus, that can be refocused by compensating the nonlinear phase at the input.

For the experimental characterization on the nonlinear TM,
the key requirement is  a medium with strong nonlinearity, and whose scattering properties can be easily controlled.
SA is an ultra-porous material made of sparse silica aggregates, 
whose optical properties can be tuned from full transparency to fully diffusive behavior, by
controlling the size and distribution of the silica inclusions \cite{kanamori2011transition}. 
Additionally, the optical properties of SA samples can be controlled by changing their spatial density, 
during fabrication \cite{jones2007method} or by mechanical compression,
e.g. for polarization or light propagation control, respectively \cite{bhupathi2010optical}.
One of the key features of SA is also its extremely low thermal conductivity, 
even lower than that of air, due to the Knundsen effect \cite{aegerter2011advances}. 
Thanks to this property, SA can host extremely high temperature spatial gradients \cite{Gentilini2014}, which in turn 
mediates optothermal nonlinearities on the order of $10^{-12} m^2/W$, 
which can be increased to $10^{-10} m^2/W$ with the inclusion of an absorbing dye \cite{braidotti2016}. 

For the SA used for this work, we adapted a standard base catalyzed silica precursor approach. Tetrametyhl orthosilicate (TMOS), Methanol, and Ammonium Hydroxide (aq) were mixed in a 2:4:1 ratio, for a total volume of 4ml.  The TMOS and Ammonium Hydroxide were purchased from Sigma Aldrich. The Ammonium Hydroxide was diluted to $35 \mu l$ per $100ml$ of ultra pure water also from Sigma Aldrich. The stock chemicals of Methanol and Acetone were purchased from Fischer Scientific. This mixture was poured in a teflon mold, producing a cuboid shaped gel, with thickness of 0.5 cm. After approximately one hour the sol-gel gelated. The newly formed gel was removed from the mould and washed in a series of several acetone baths, each lasting 24 h, to remove residual chemical impurities left over from the gelation process. A few drops of water were added in the final acetone bath, to add a small impurity that would lead to a small amount of opaqueness to the final sample, to increase the amount of scattered photons. The aerogel were dried using a custom made ${\rm CO}_2$ critical point drier, over a period of 24 h.

For the nonlinear experiments, we used a pump-probe setup coupled with a spatial light modulator (SLM) for wavefront shaping of the probe beam. The setup, as well as details on the linear absorption of the SA at the pump and probe wavelength can be found in figures S1 and S2 of the SI.

\begin{figure}[htbp]
\centering
\includegraphics[width=8.6 cm]{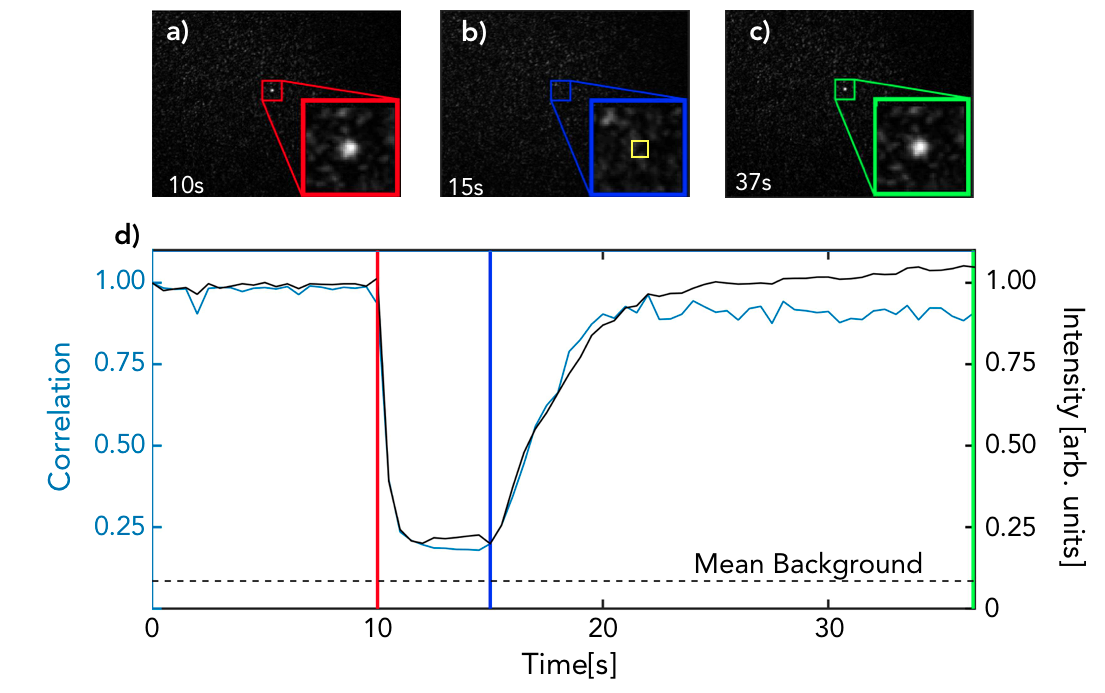}
\caption{a-c) CCD images of the output facet of the sample at different instants. The inset are zoomed in views around the Region of Interest (ROI), which is a 8x8 pixel region, highlighted in yellow in panel b. Panel d) shows the intensity in the Region of Interest (ROI) and the cross correlation of the entire speckle pattern over time with respect to the t=0 pattern.}
\label{fig2}
\end{figure}

To demonstrate the dynamic control of the transmission channels, we first adopted a wavefront shaping approach. 
The probe light was focused in the SA in a 8x8 pixel wide region of interest (ROI) on the CCD display, by employing a genetic algorithm \cite{conkey2012genetic}. 
With this approach we found an enhancement of the intensity in the ROI of 12.5x with respect to the background intensity. This is comparable with reported values in other materials and algorithms \cite{Vellekoop2008}. The focused spot was controlled reversibly switching on and off the pump, with pump power equal to 200mW, as seen in \ref{fig2}a-c. 

The focusing through random media via wavefront shaping is enabled by long-range correlations between the channels \cite{rotter2017light}.
Pumping dephases the channels and eliminates the bright point. 
Several processes  can lead to this dephasing: absorption \cite{liew2015modification}, refractive index changes, and scattering particle deformation \cite{abb2011ultrafast}. 
Absorption was eliminated by a direct transmission measurement, 
observing that no attenuation occurs at the probe wavelength when the sample is pumped, as shown in the SI. 
The other two mechanisms are interlinked and can be treated together. 
SA is a highly porous skeletal structure that exhibits a high degree of optothermal nonlinearity. 
Thermal effects due to pumping result in an expansion of the silica matrix, lowering its density, resulting in a refractive index change, which dephases light. 

As seen in fig. \ref{fig2}d, when pumping the SA, the intensity drop in the ROI was approximately 80\% compared to its peak value.
Once the pump was turned off and the material cooled down, the focus point returned to the same intensity it was prior to the pumping process. 
Remarkably, the cross-correlation of the whole image with a pre-pumped reference frame was fully restored shortly after the pump was switched off.
Both these effects demonstrate that the nonlinear deformation of the SA skeletal framework is a fully reversible process.  
The minimum value of the cross-correlation was similar to that obtained correlating the reference frame with a random, uncorrelated speckle distribution.

We next set out to measure the TM of the SA, under different pumping conditions. 
The phase front of the incoming beam was manipulated using a sequential algorithm, 
in order to determine contributions from individual input channels.  
The 512x512 pixels of the SLM were grouped in a 16x16 array, which optimized the signal to noise ratio. 
The phase of each pixels was tuned from 0 to 2$\pi$. 
For each pixel value we acquired 1080x1080 pixels images. To improve the image stability, 
we integrated the signal in 120x120 pixels images and averaged them over 10 acquisitions. 

\begin{figure}[htbp]
\centering
\includegraphics[width=8.6 cm]{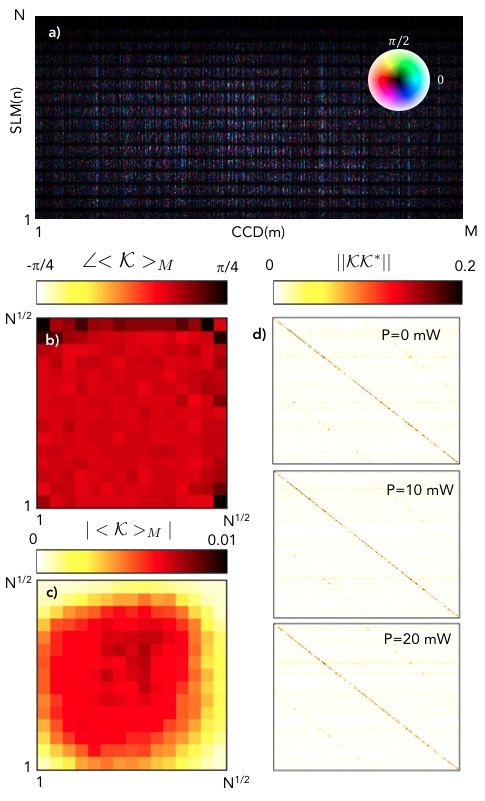}
\caption{a) Amplitude (brightness) and phase (color) of a typical TM. b-c) Phase and amplitude of the TM, averaged over the M pixels of the CCD and remapped on 16x16 SLM matrices, respectively.  d) 256x256 pixel wide section of the norm of a TM multiplied by its normalized complex conjugate.}
\label{fig3}
\end{figure}

The amplitude of each element of the TM is given by the peak to peak value of the intensity recorded in a CCD pixel, as the phase of the relative SLM pixel is varied. 
The phase of each element is given by the offset between the maximum intensity and the phase of the SLM pixel. This procedure is discussed in details in the SI.
The resulting TM comprises 256x14400 elements and the complex coefficients of a typical TM are shown in Fig. \ref{fig3}a. 
To facilitate the analysis of the results, it is useful to take the average of the TM over the M pixels of the CCD images and plot its amplitude and phase, as shown in Fig. \ref{fig3}b-c, 
remapped on the 16x16 coordinates of the SLM.
From a physical point of view, this corresponds in taking the average effect that a single pixel of the SLM (which corresponds to a specific direction of light incident on the sample) has on the brightness of the image acquired by the CCD. 
The measurement process was repeated at a range of pump powers from 5mW to 20mW. 
In each case the SA was pumped for 5 minutes before measurement after which there was no change in the speckle pattern.
To generate a high intensity on the $i^{th}$ pixel of the CCD, it is sufficient to consider the $i^{th}$ row of the phase of the TM and remap it to the coordinate space of the SLM. 
An indication of the quality of the TM is obtained by multiplying each matrix by its normalized complex conjugate which, in the ideal case, results in a diagonal matrix \cite{popoff2010}.
Three representative examples are shown in fig. \ref{fig3}d, where we selected a central area of 256x256 elements of the norm of the full 14400x14400 matrix.

\begin{figure}[htbp]
\centering
\includegraphics[width=8.6 cm]{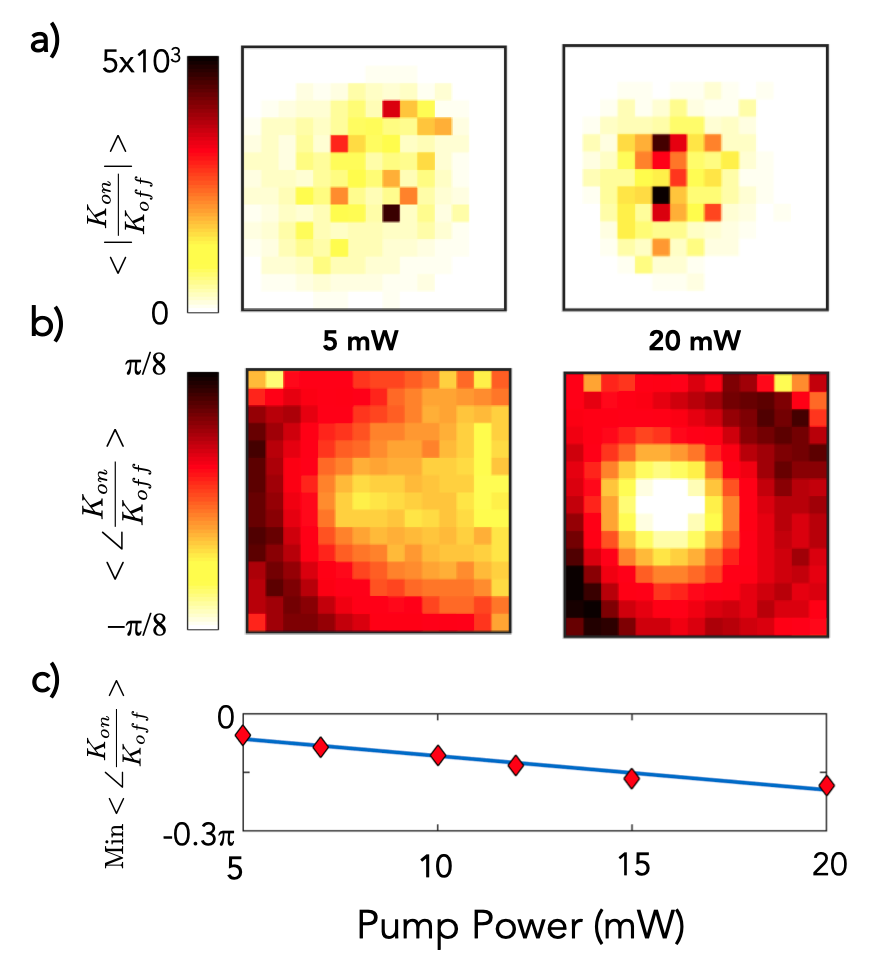}
\caption{
Measurement of the nonlinear transmission matrix: (a) ratio between the amplitudes of the TM at different powers with $K_{off}$;
(b) nonlinear phase at different powers determined as the phase of the
ratio between $K_{on}$ and $K_{off}$;¤
(c) Amplitude of the peak of the dephasing vs pump power, extracted from panel b. }
\label{fig4}
\end{figure}

To determine the relationship between the TM obtained at different pumping conditions, as shown in fig. \ref{fig4}a-b, 
we took the amplitude and phase of the ratio between the TMs of the pumped and un-pumped sample, element by element, averaged along the CCD coordinates and remapped in the SLM space, as for fig. \ref{fig3}.
The specific profiles emerging in the differential matrix are due to the scattering properties of the sample, to the particular choice of basis used to address the SLM and on the fact that we slightly overfilled the back focal plane of the focusing lens. This is confirmed by the fact that the amplitude of the unperturbed TM has the same shape (see fig. \ref{fig3}c).
In this configuration, the photons reflected from the external pixels of the SLM do not couple to propagating channels in the medium. 
This effect is particularly evident when considering the perturbation of the TM, both in amplitude and phase. 
Additionally, the sign of the dephasing is consistent with the fact that the optothermal nonlinearity in the SA is self defocusing \cite{braidotti2016}.

The peak phase change in the TMs for different pump powers is shown in fig. \ref{fig4}c. The amplitude of the maximum dephasing increases linearly up to 20mW. 
After this pump value, the pumped material changes substantially and the transmission channels become uncorrelated with the unpumped case. 
This trend demonstrates that for low pump powers the channels are only slightly dephased, until the deformation of the material induces a completely different scattering dynamics, embodied by an uncorrelated set of transmission channels. 
In order to restore the focusing action, the SLM must compensate this nonlinear phase perturbation. And this is exactly what we found from the experiments.

The transmission matrices formalism for random media can usually be used either to engineer an input complex field to obtain a pre-determined output field or to infer the input field, from the analysis of an output image \cite{rotter2017light}.  With our result we extend the applicability of the TM paradigm to nonlinear samples with all-optical control of the material properties, for known pumping conditions. Additionally, for our specific case, there exists a pumping regime where the evolution of the channels is correlated, thus allowing to use the TM predictively.
We anticipate that the use of the TM in the nonlinear regime will enable a new class of nonlinear devices based on all-optical control random media, e.g. in random lasing, imaging, gating and switching applications.  

%\section*{Conclusion}
We have demonstrated the reversible and all-optical control of wavefront shaping in nonlinear random media based on opto-thermal nonlinearity in silica based aerogel. We also characterized the transmission matrix of the random material in the nonlinear regime, using a pump and probe scheme coupled with phase controlled input modulation. In the material under examination we identified a regime where the transmission channels maintain a degree of correlation for a finite range of powers. The full knowledge of TM in the nonlinear regime paves the way for a new generation of nonlinear devices based on random media.

\subsection*{Acknowledgments}
This project was supported by Sapienza Visiting scholarship scheme. ADF and AF thank EPSRC (EP/M508214/1). CC acknowledges QuantERA Quomplex (grant number 731743).

%\subsection*{Authors contributions}
%AF prepared the sample and performed the experiments. CC developed the theoretical interpretation. ADF directed the work. All authors analyzed the data and prepared the article.

%\bibliography{SLM}


\end{document}

\title{Nonlinear transmission matrices of random optical media - Supplementary Information}
 \author{A. Fleming$^1$, C. Conti$^{2\dag}$, A. Di Falco$^1$} 
\email{adf10@st-andrews.ac.uk, \dag claudio.conti@uniroma1.it} 
%\thanks{Corresponding author}
\affiliation{
$^1$School of Physics and Astronomy, University of St. Andrews, North Haugh, St. Andrews KY16 9SS, UK\\
$^2$ Department of Physics, University Sapienza, Piazzale Aldo Moro 2, 00185, Rome, Italy}

%\date{\today} 
\maketitle
 
%\pacs{05.45.Mt, 42.25.-p, 05.40.-a} 
\section*{S1 - Pump-Probe Optical Setup}

The probe beam ($\lambda_{probe}=830~\rm{nm}$, MDL-III-830-800mW diode from Changchen New Industries Optoelectronics Technology Co., Ltd.) was polarized and collimated onto a spatial light modulator (SLM) (HSP512 from Boulder Nonlinear Systems). 
The image displayed on the SLM was relayed on the back aperture of a 30x ashperic lens ($f=6.2mm$), to focus the light on the sample. 
This configuration allowed correlating the change in phase of the pixels of the SLM to the direction of light impinging onto the sample \cite{popoff2010}.
The scattered light was collected by a 20x objective from Newport ($f=9.0mm$, NA$=0.40$), with a field of view of $400~\rm{\mu m}$ in diameter, and imaged onto a CCD camera (Basler acA1920-25gm). 
To ensure the collection of only scattered photons, two cross polarizers were used on either side of the sample, with a measured fraction of collected light of 26\%.
The pump beam ($\lambda_{pump}=488~\rm{nm}$)  was focused on the back focal plane of the input objective, 
creating a collimated beam $40\rm{\mu m}$ in width, collinear to the probe beam. 

\begin{figure*}[htbp]
\centering
\includegraphics[width=.7\linewidth]{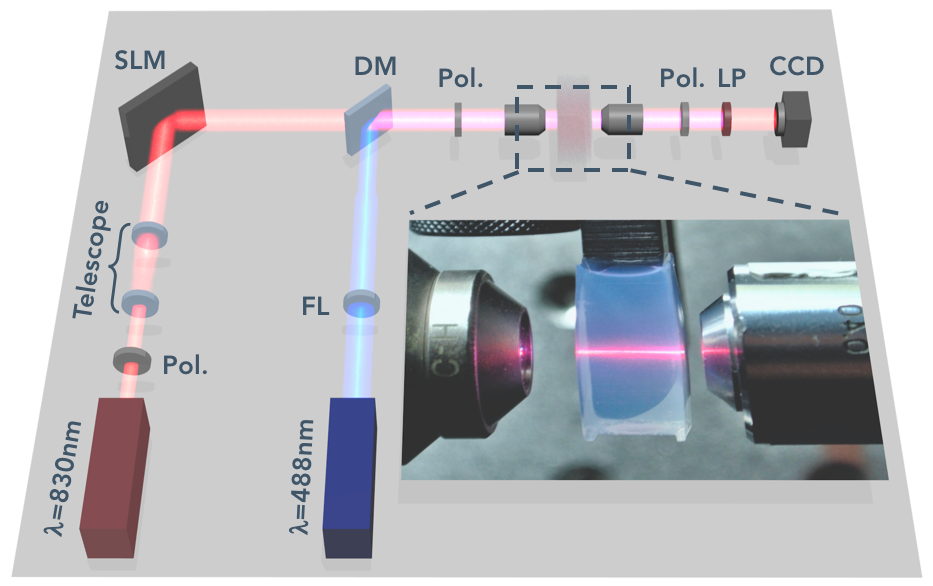}
\caption{ Pump-Probe Optical setup with wavefront shaping of the probe beam by SLM.}
\label{figS1}
\end{figure*}

\section*{S2 - linear absorption of Silica Aerogel}

\begin{figure*}[htbp]
\centering
\includegraphics[width=.7\linewidth]{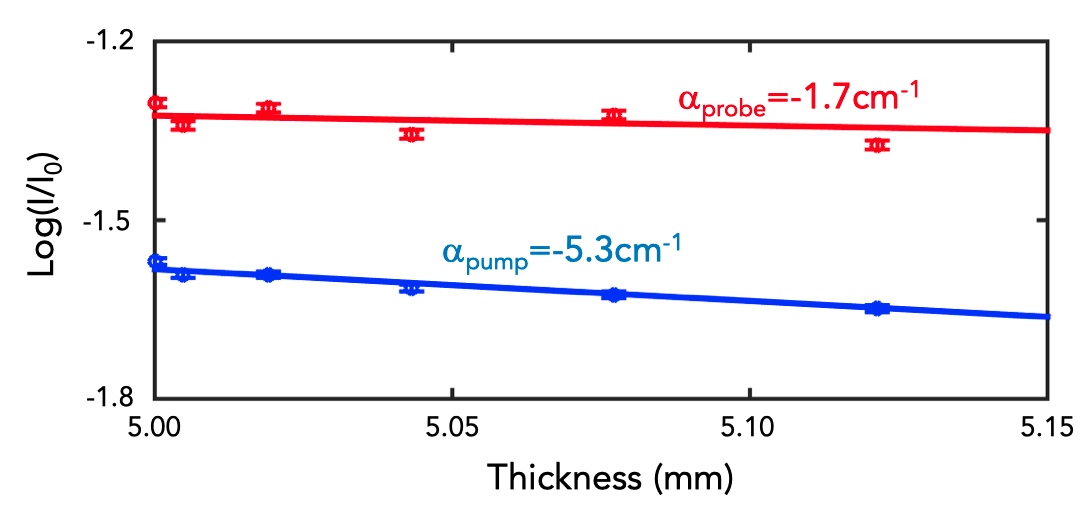}
\caption{Transmission characteristics of the SA for the pump and probe wavelengths.}
\label{figS2}
\end{figure*}

The linear absorption of the used SA sample was estimated measuring 
the optical transmission of the sample for different angles and unpolarized, collimated light, and shown in fig. S\ref{figS2}.

From the values of $\alpha$ it is possible to evaluate the scattering mean free path ($l_{pump}=1/\alpha_{pump}=1.9$~mm; $l_{probe}=1/\alpha_{probe}=5.9$~mm) and transport mean free path ($t_{pump}=l_{pump}/(1-g) =2.1$~mm; $t_{probe}=l_{probe}/(1-g)=6.5$~mm), where for the directionality factor we assumed the value $g=0.1$, as typical of silica aerogel samples \cite{hulst1981light}. Therefore we can conclude that the experiments were completed in the weakly scattering regime.

\begin{figure*}[htbp]
\centering
\includegraphics[width=.7\linewidth]{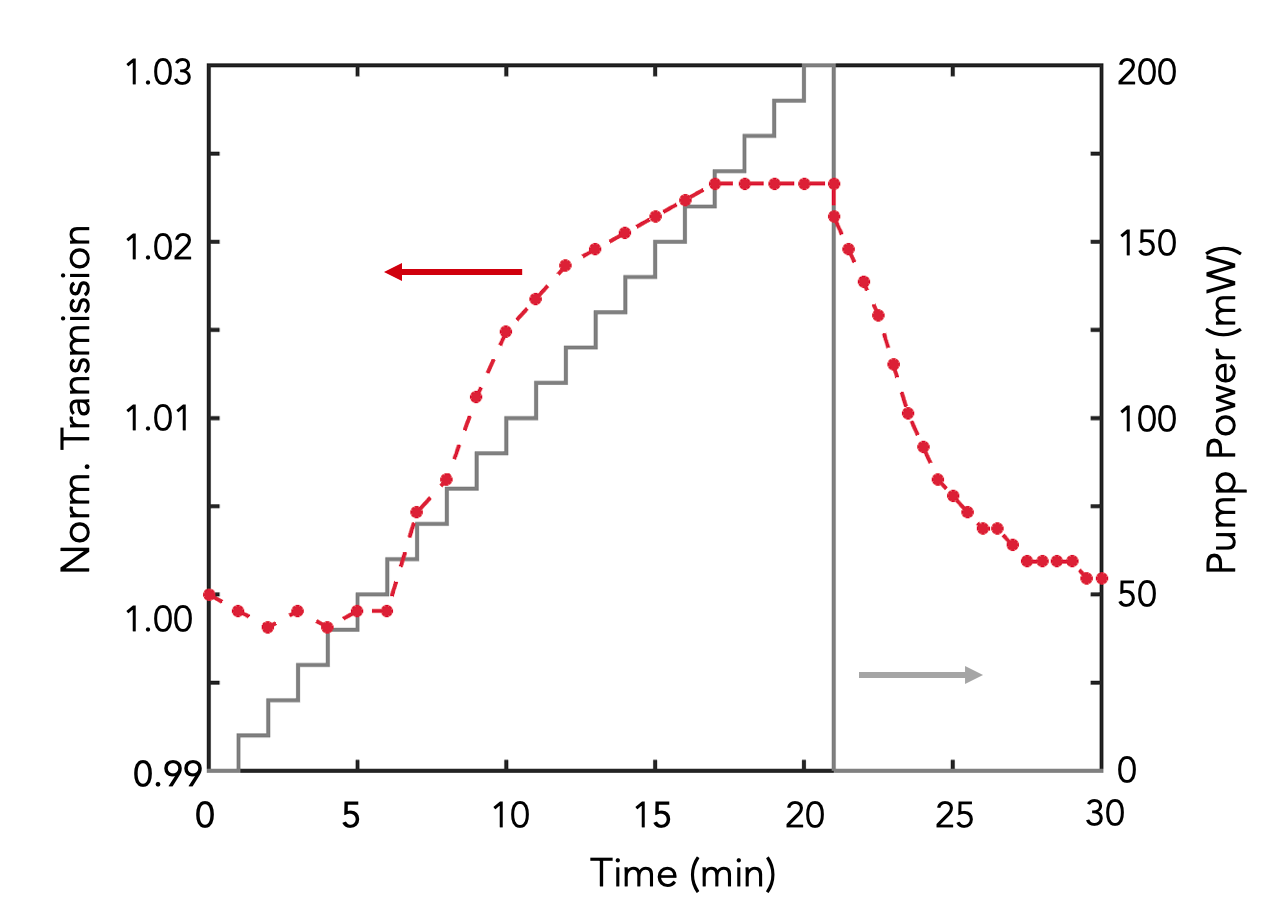}
\caption{Transmission characteristics of the SA at the probe wavelength (left axis) and pump power (right axis) vs time.}
\label{figS3new}
\end{figure*}

To exclude pump induced absorption of the aerogel, we characterized the transmission of the sample at the probe wavelength vs time, for different values of the pump. Fig. S\ref{figS3new} 
shows that the normalized transmission for a collimated probe increases marginally when the sample is collinearly pumped. This is in keeping with the fact that the aerogel has a defocusing nonlinearity, therefore it becomes slightly less dense, and thus less scattering medium.

\section*{S3 - Construction of the transmission matrices}

\begin{figure*}[htbp]
\centering
\includegraphics[width=0.8\linewidth]{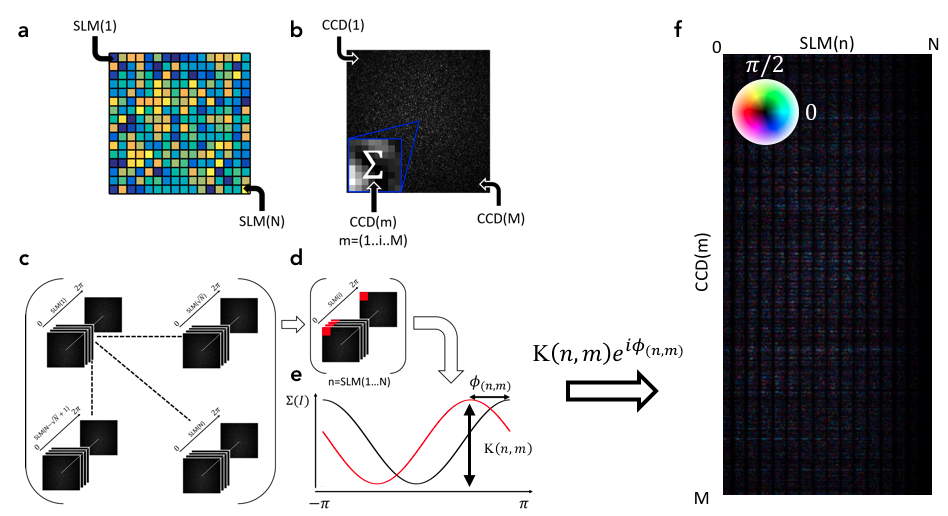}
\caption{Process outline for the determination of the Complex Transmission Matrices.}
\label{figS3}
\end{figure*}

The process for forming the TM from raw image data is outlined in figure S\ref{figS3}. The 2D pixels of the CCD (M pixels) and of the SLM (N pixels) are mapped in a MxN TM matrix. 
To improve the SNR in the CCD images, we sum the total black-white intensity values over 8x8 pixels, giving a measurement range between 0 and 16383, rather than 0 to 255.

The phase of each pixel of the SLM is tuned in turn in the range $(-\pi,\pi)$, keeping the other pixels at $-\pi$ and the corresponding CCD image is acquired. The light impinging on the constant area of the SLM interferes with that of the tuned pixel, to access the complex values of the transmission channel. This process produces a stack of 3D images for each SLM pixel, as shown in panel c).

The intensity of each pixel in the stack changes with the phase of the SLM pixel in a cosine function. The amplitude and phase of the relative elements of the TM are given by the peak-to-peak value of the cosine function and by the offset respect to the reference phase, respectively, as seen in panels d-e). A typical complex TM is shown in panel f).

\section*{S4 - Transmission matrices in the nonlinear regime}
To model the transfer matrix in the presence of an external perturbation, it is convenient to use a Green function formalism \cite{economou,rotter2017light}.
Following this approach, the field distribution in a scattering medium can be described by $|{\bf E}\rangle={\bf K}|{\bf E}_0\rangle$,
where ${\bf E}_0$ is the incident field and ${\bf K}={\bf 1}-{\bf Ge}_s$ is a generalized propagator, where {\bf 1} is the unitary matrix and the Green function {\bf G} is such that
  \begin{equation}
    \label{eq:10}
\left(\bm{\mathcal{D}}+{\bf e}\right){\bf G}={\bf 1}.
  \end{equation}

Here, $\bm{\mathcal{D}}(\mbr)=-\nabla\times\nabla\times$ and ${\bf e}={\bf e}_b+{\bf e}_s$ is the operator  
\begin{equation}
  \label{eq:11}
  \langle \mbr | {\bf e} |\mbfr'\rangle=k_0^2 \varepsilon(\mbr)\delta(\mbr-\mbr')
\end{equation}  
  
 associated to the relative permittivity $\varepsilon({\bf r})=\varepsilon_b({\bf r})+\varepsilon_s({\bf r})$, where $\varepsilon_b({\bf r})=1$ is the permittivity of the homogenous background medium and $\varepsilon_s({\bf r})$ is the permittivity of the scattering medium.
 
In position ${\bf r}$ representation the propagator can be written as 
  \begin{equation}
    \label{eq:30}
    \langle \mbr| {\bf K}| \mbr' \rangle={\bf 1}\delta(\mbr-\mbr')-
    k_0^2 \varepsilon(\mbr')\langle \mbr|{\bf G}| \mbr'\rangle .
  \end{equation}
  and its matrix elements are
  \begin{equation}
    k_{mn}=\langle m | {\bf K} |  n \rangle.
    \end{equation}
 In the presence of the perturbation due to the pumping, the perturbed propagator is
  \begin{equation}
    \label{eq:30p}
    {\bf K}'={\bf 1}-{\bf G}'{\bf e}'
  \end{equation}
with ${\bf G}'$ the perturbed Green's function such that 
   \begin{equation}
    \label{eq:26d}
    \left(\bm{\mathcal{D}}+{\bf e}_b+{\bf e}_s+{\bf e}'\right){\bf G}'={\bf 1},
  \end{equation}
 and ${\bf e}'$ is the operator associated to the perturbed permittivity $\Delta\varepsilon({\bf r})$, where $\varepsilon(\mbr)=\varepsilon_b(\mbr)+\varepsilon_s(\mbr)+\Delta \varepsilon(\mbr)$.
The state in the presence of perturbation $|{\bf E}'\rangle$ can then be expressed in terms
  of the state without perturbation  $|{\bf E}\rangle$
  and the input state $|{\bf E}_0 \rangle$
  as operator multiplication
  \begin{equation}
    \label{eq:29}
    |{\bf E}'\rangle={\bf K}'|{\bf E}\rangle=
    {\bf K}'{\bf K}|{\bf E}_0\rangle.
  \end{equation}
 Correspondingly, the transmission matrix elements $k_{mn}^{\text{NL}}$ in the presence of the nonlinear perturbation can be written as a matrix multiplication
  \begin{equation}
    k_{mn}^{\text{NL}}=k'_{mq}k_{qn}
  \end{equation}
  where we omitted the sum over the repeated symbol $q$.
  By using (\ref{eq:30p}), the element of the rotation matrix $k_{mq}'$ is written as
  \begin{equation}
    k'_{mq}=
    \delta_{mq}+w_{mq},
  \end{equation}
  with $\delta_{mq}$ the Kronecker symbol and the perturbation elements
  \begin{equation}
  \label{eq:wmq} 
    w_{mq}=-\langle m | {\bf G}'{\bf e}'| n\rangle.
  \end{equation}

The element of the perturbed matrix can then be written as 
  \begin{equation}
    k_{mn}^{\text{NL}}=k_{mn}+w_{mq}k_{qn}=k_{mn}+w_{m1}k_{1n}+...+w_{mN}k_{Nn}.    
    \label{eq:kprime}
  \end{equation}
  
  Eq.~(\ref{eq:kprime}) can be interpreted as follows:
  in the absence of perturbation light is channelled - with amplitude proportional to $k_{mn}$ - from the channel $n$ to the channel $m$; in the presence of the perturbation,
  further contributions arise from other channels. For example, the light
  channeled from $n$ to $1$ with amplitude $k_{1n}$ also contributes
  to the signal in the channel $m$ with amplitude $w_{m1}$.
  This may be described by stating that nonlinearity add furthers
  channels for light by scattering from one unperturbed channel to another.

  Eq.~(\ref{eq:kprime}) can be written following \cite{vellekoop2008phase}:
  \begin{equation}
    k_{mn}^{\text{NL}}=k_{mn}\frac{1+\xi_{mn}}{\sqrt{1+2\phi_{NL}^2}},
   = k_{mn}e^{\imath \kappa_{mn}\phi_{NL} }   
    \label{eq:kprime1}
  \end{equation}
  being $\xi_{mn}$ a complex Gaussian variable with zero mean and (for small perturbations $\phi_{NL}$) $\langle |\xi_{mn}|^2 \rangle=2\phi_{NL}^2$,
  and defining the modal dependent coefficients by
  \begin{equation}
    \kappa_{mn}\phi_{NL}= \arg(1+{\xi_{mn}})\simeq\Im({\xi_{mn}})
    \label{kappaphase}
    \end{equation}
    such that $\phi_{NL}$ represents the average phase shift of the mode,
    and $\langle |\kappa_{mn}|^2 \rangle=1$.
Additionally, as the overall transmission of the sample changes in a negligible way, the transmission matrix is such that 
\begin{equation}  
  \langle |k_{mn}^{\text{NL}}|^2 \rangle= \langle |k_{mn}|^2\rangle.    
\label{eq:kmnnl}
\end{equation}  

\noindent{\bf Theoretical estimate of the perturbation ---}
To obtain a theoretical estimate of the parameter $\phi_{NL}$ we make use of eqs. (\ref{eq:kprime}) and (\ref{eq:kprime1}) to show that
  \begin{equation}
    2\phi_{NL}^2=\langle |\xi_{mn}|^2 \rangle=
    \langle |\sum_q \frac{w_{mq} k_{qn}}{k_{mn}}|^2 \rangle= \langle |\sum_q w_{mq}|^2 \rangle=2\langle \left[\Im\left(\sum_q w_{mq}\right) \right]^2 \rangle,
    \label{phi1}
  \end{equation}

which means that $\phi_{NL}$ represents the standard deviation of a Gaussian variable (the sum of many complex variables), which is independent of the mode indices $m$ and $n$ as is true for the average of $k_{mn}$.
Therefore, the bracket in (\ref{phi1}) can be taken as average
  over the modes and the disorder realizations.

From eq. (\ref{eq:wmq}) we have 
  \begin{equation}
    \sum_q w_{mq}=\sum_q \langle m | {\bf G}'{\bf e}'| q\rangle
    \simeq \sum_q\langle m | {\bf G}{\bf e}'| q\rangle 
  \end{equation}

  Where we have used ${\bf G}'\simeq {\bf G}$ as we are interested
  in the lowest order approximation with respect to ${\bf e}'$.
 
  By using the modal representation of the Green function \cite{economou}
\begin{equation}
  \label{eq:15b}
  {\bf G}=c^2 \sum_j \frac{|j\rangle \langle j|}
  {\omega^2-\omega_j^2}\text{,}
\end{equation}
where we adopt the canonical orthonormal set, gives
  \begin{equation}
    \sum_qw_{mq}=\sum_q \frac{ \langle m | c^2 {\bf e}'| q\rangle}{\omega^2 -\omega_m^2}=\sum_q  \frac{ \omega^2}{\omega^2 -\omega_m^2}
    \int \Delta\varepsilon(\mbr) \bm{\phi}_m(\mbr)^*\cdot \bm{\phi}_q(\mbr)\D \mbr
  \label{eq:sumb}    
  \end{equation}
  where in the last equation we used the position representation.
  
  A further simplification  can be obtained by observing that eq. (\ref{eq:sumb})
  is the sum of $N$ terms which all are of the order of
  $\int \Delta\varepsilon(\mbr) \bm{\phi}_m(\mbr)^*\cdot \bm{\phi}_m(\mbr)\D \mbr$
  if $\Delta\varepsilon(\mbr)$ is a perturbation that involves most of the
  sample and couples all the modes, and if the modes are not strongly localized. 
  In this approximation we can write
  \begin{equation}
    \sum_qw_{mq}\simeq  N \frac{ \omega^2}{\omega^2 -\omega_m^2}
    \int \Delta\varepsilon(\mbr) \bm{\phi}_m(\mbr)^*\cdot \bm{\phi}_m(\mbr)\D \mbr
  \label{eq:sumc}    
  \end{equation}
  
    We recall one can write
  \begin{equation}
    \frac{1}{\omega^2-\omega_m^2}=  PV\left[\frac{1}{\omega^2-\omega_m^2} \right]+\frac{\imath \pi}{2\omega}\delta(\omega-\omega_m)
    \label{PVequation}
  \end{equation}
with $PV$ the principal value. As $\xi_{mn}$ is the sum of many random contributions, the real and the imaginary part will be Gaussian variables with the same
variance.

Averaging (\ref{eq:sumc}) over all the modes (the average quantities are expected to be modal independent), by summing w.r.t. to the index $m$ and dividing by $N$ we have
  \begin{equation}
   \Im \left(\sum_qw_{mq}\right)\simeq  \frac{ \pi \omega}{2}
    \int \Delta\varepsilon(\mbr) \rho(\mbr,\omega)\D \mbr
  \label{eq:sume}    
  \end{equation}
where we used the expression for the LDOS
\begin{equation}
\rho(\mbr,\omega)=\sum_m \delta(\omega-\omega_m)\bm{\phi}_m(\mbr)^*\cdot \bm{\phi}_m(\mbr) 
\end{equation}
  
Finally we have
\begin{equation}
  \phi_{NL}^2 =\frac{1}{2}\langle |\sum w_{mq}|^2 \rangle=
  \langle\left[\Im \left(\sum_q w_{mq}\right)\right]^2 \rangle  \simeq  \frac{\pi^2 \omega^2}{4}\langle \left(\int \Delta\varepsilon(\mbr) \rho(\mbr,\omega)\D \mbr \right)^2\rangle.      
\end{equation}